\documentclass{PoS}

\title{Search for extragalactic astrophysical counterparts of IceCube
   neutrino events}

\ShortTitle{IceCube neutrino source}

\author{\speaker{Reetanjali Moharana}\\
        Dept.\ of Physics, University of Johannesburg, PO Box 524, Auckland Park 2006, South Africa \\
        E-mail: \email{reetanjalim@uj.ac.za}}
        
\author{{Richard J.G. Britto}\\
        Dept.\ of Physics, University of Johannesburg, PO Box 524, Auckland Park 2006, South Africa 
        }                
        
\author{{Soebur Razzaque}\\
        Dept.\ of Physics, University of Johannesburg, PO Box 524, Auckland Park 2006, South Africa \\
        E-mail: \email{srazzaque@uj.ac.za}}

\abstract{Detection of 54 very high-energy (VHE) neutrinos by the IceCube Neutrino Observatory has opened a new chapter in multi-messenger astronomy. However due to large errors in measuring the directions of the neutrino shower-type events, which dominate the current event list, it is difficult to identify their astrophysical sources. We perform cross-correlation study of IceCube neutrino events with extragalactic candidate sources using X-ray and gamma-ray selected source catalogues such as Swift-BAT, 3LAC and TeV-Cat. We apply different cuts on the X-ray and gamma-ray fluxes of the sources in these catalogs, and use different source classes in order to study correlation. We use invariant statistic and Monte Carlo simulations to evaluate statistical significance of any correlation.}

\FullConference{The 34th International Cosmic Ray Conference,\\
		30 July- 6 August, 2015\\
		The Hague, The Netherlands}

\begin{document}

\section{Introduction}
IceCube Neutrino Observatory, the world's largest neutrino
detector, has detected 54  neutrino events within 1347 days with energy between 20 TeV to 2.3 PeV ~\cite{aartsen2014observation,kop}. 
Shower events, most likely due to $\nu_e$ or
$\nu_\tau$ charge current $\nu N$ interactions and also due to neutral current $\nu N$ interactions of all flavors, dominate the event
list (39 including 3 events with 1--2 PeV energy) while track events,
most likely due to $\nu_\mu$ charge current $\nu N$ interactions,
constitute the rest. Among a total of 54 events about 21 could be due
to atmospheric neutrino ($9.0^{+8.0}_{-2.2}$) and muon ($12.6\pm 5.1$)
backgrounds.  A background-only origin of all 54 events has been
rejected at 6.5-$\sigma$ level ~\cite{kop}. Therefore a
cosmic origin of a number of neutrino events is robust. The track
events have on average $\sim 1^\circ$ angular resolution, but the
dominant, shower events have much poorer angular resolution, $\sim
15^\circ$ on average ~\cite{kop}. Searching for sources of these events is now one of the major
challenges in astrophysics. Pinpointing the astrophysical sources
where these neutrinos are coming from is difficult due to large
uncertainty in their arrival directions.

High energy cosmic rays (CRs)
can interact with low energy photons and/or low energy protons to
produce neutrinos and high energy gamma rays inside the source or
while propagating to earth. So a multi-messenger study of neutrinos, Cosmic Rays (CRs)
and gamma-rays can identify the possible astrophysical sources. In our first attempt to search for sources we tried to see a correlation with Ultra-High Energy (UHE) CRs with the earlier 37 cosmic neutrino events ~\cite{Moharana:2015nxa}. A detail analysis of IceCube neutrino events with the Pierre Auger Observa (PAO) and Telescope Array (TA) has been done in collaboration ~\cite{Aartsen:2015dml}.

Here we study correlation of IceCube neutrino events with TeVCat, {\it Swift}-BAT 70 month X-ray source catalog ~\cite{2013ApJS..207...19B} and 3LAC source catalog ~\cite{Ackermann:2015yfk}. A similar study of correlation of IceCube neutrinos with the gamma ray sources has also been done ~\cite{Resconi:2015nva} to find a correlation of $Fermi$-LAT source with only track HESE events is also done ~\cite{anthony}. Recently a detail analysis of correlation showed at least 2 $\sigma$ result with extreme blazars ~\cite{Resconi2} and 3$\sigma$ with the starforming regions ~\cite{Emig:2015dma}. To do specific correlation study we use different cuts on the energy flux of these sources, and also different sets of source types, and showed the results of this study.

\begin{figure}[h]
\includegraphics[width=36pc]{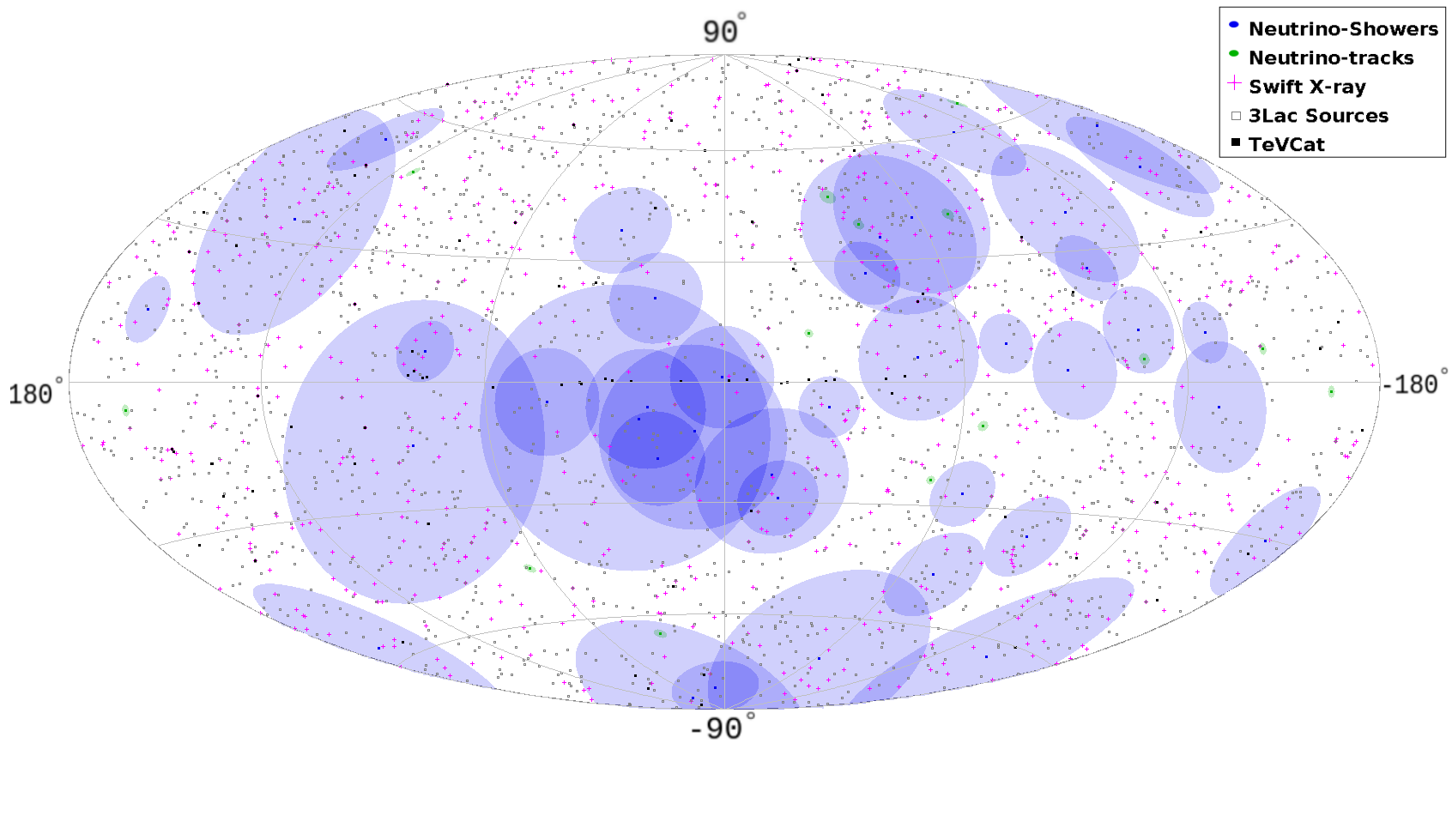}
\caption{\label{skymap}Sky map of the 52 IceCube cosmic neutrino events with error circles and sources from different catalogs in Galactic coordinate system.}
\end{figure}

\section{IceCube neutrino events and Source catalogs}

For our analysis we consider all 52 IceCube detected neutrino events. Two track events (event numbers 28
and 32) are coincident hits in the IceTop surface array and are almost
certainly a pair of atmospheric muon background events
~\cite{aartsen2014observation}. Therefore we excluded them from our analysis.
Fig.~\ref{skymap} shows sky map of the 52  events in Galactic 
coordinates with reported angular errors.

For the correlation analysis we have used 3 different source catalogs. 
{\it Swift}-BAT 70 month X-ray source catalog ~\cite{2013ApJS..207...19B}, $Fermi$ Third Catalog of Active Galactic Nuclei (3LAC) ~\cite{Ackermann:2015yfk}, TeVCat ~\cite{2008ICRC....3.1341W}. The sky map in Fig.~\ref{skymap} shows the extragalactic sources from these catalogs.

{\it Swift}-BAT 70 month X-ray source catalog includes 1210 objects, and after excluding Galactic sources the number of sources become 785. In our previous study ~\cite{Moharana:2015nxa} we found 18 sources from this catalog that are correlated simultaneously with UHECRs and IceCube neutrino events. PAO collaboration has also found an anisotropy at $\sim 98.6\%$ CL in
UHECRs with energy $\ge 58$ EeV and within $\sim 18^\circ$ circles
around the AGNs in {\em Swift}-BAT catalog at
distance $\le 130$ Mpc ~\cite{PierreAuger:2014yba} . These 18 sources mostly have an X-ray energy flux $\ge10 ^{-11}$ ${\rm{erg} \, \rm{cm^{-2}} \, \rm{sec}^ {-1}}$. So, in the present analysis we use all the sources from this catalog which have flux $\ge10 ^{-11}$ ${\rm{erg} \, \rm{cm^{-2}} \, \rm{sec}^ {-1}}$. This condition decreased the number of sources to 687. In the sky map of Fig.~\ref{skymap} we have shown these 687 sources. 

TeVCat contains sources that are detected with very high energy (VHE) gamma rays with energy $\ge 50$ GeV. It includes 161 sources, out of which 22 are unidentified sources. This is the highest energy source catalog, particularly interesting for $\nu$ production. Sky map in Fig~\ref{skymap} contains TeVCat sources that are not in the Galactic plane.

The {\it Third Catalog of Active Galactic Nuclei (AGNs)} detected by Fermi LAT (3LAC) ~\cite{Ackermann:2015yfk} is a subset of the {\it Fermi} LAT {\it Third Source Catalog (3FGL)} ~\cite{2015ApJS..218...23A}. The 3FGL catalog includes 3033 sources detected above a 4$\sigma$ significance (test statistic $>$ 25) on the whole sky, during the first 4 years of the Fermi mission (2008-2012). The original 3LAC sample includes 1591 AGNs from 3FGL, though 28 are duplicate associations. An additional cut had also been performed to exclude the Galactic plane region ($|b| \leq  10^\circ$) where the incompleteness of the counterpart catalogs significantly hinders AGN association. However, in this paper, we chose to study what we call the ``extended 3LAC" sample of 1773 sources, that includes sources of the Galactic plane, and that could be associated to several neutrino events. In the extended 3LAC sample, 491 sources are flat spectrum radio quasars (FSRQs), 662 are BL Lacs, 585 are blazars of unknown type (BCU), and 35 are non-blazar AGNs. 

\section{Statistical method for Correlation study}
To study correlation between cosmic neutrinos and sources from different catalogs separately, we map the
Right Ascension and Declination $(RA, Dec)$ of the event directions and sources
into unit vectors on a sphere as
$$
{\hat x} = (\sin\theta \cos\phi, \sin\theta \sin\phi, \cos\theta)^T,
$$
where $\phi = RA$ and $\theta = \pi/2 - Dec$. Scalar product of the
neutrino and source vectors $({\hat x}_{\rm neutrino}\cdot {\hat
  x}_{\rm source})$ therefore is independent of the coordinate system.
The angle between the two vectors
\begin{equation}
\label{gamma}
\gamma = \cos^{-1} ({\hat x}_{\rm neutrino}\cdot 
{\hat x}_{\rm source}),
\end{equation}
is an invariant measure of the angular correlation between the
neutrino event and source directions ~\cite{Virmani:2002xk,Moharana:2015nxa}. Following ref.~\cite{Virmani:2002xk} we use a
statistic made from invariant $\gamma$ for each neutrino direction
${\hat x}_i$ and source direction ${\hat x}_j$ pair as
\begin{equation}
\label{delta}
\delta\chi^2_i = {\rm min}_j (\gamma_{ij}^2/\delta\gamma_i^2),
\end{equation}
which is minimized for all $j$. Here $\delta\gamma_i$ is the
1-$\sigma$ angular resolution of the neutrino events. We use the exact
resolutions reported by the IceCube collaboration for each event
~\cite{aartsen2014observation}.

A value $\delta \chi^2_i \le 1$ is considered a ``good match'' between
the $i$-th neutrino and a source directions. We exploit
distributions of all $\delta\chi^2_i$ statistics to study angular
correlation between IceCube neutrino events and sources in catalog.  The
distribution with observed data giving a number of ``hits'' or $N_{\rm
  hits}$ with $\delta\chi^2 \le 1$ therefore forms a basis to claim
correlation. Note that in case more than one source direction from the catalog are
within the error circle of a neutrino event, the $\delta\chi^2$ value
for UHECR closest to the neutrino direction is chosen in this method.

We estimate the significance of any correlation in data by comparing
$N_{\rm hits}$ with corresponding number from null distributions.
We construct null distributions by randomizing only the $RA$ of the sources, keeping their $Dec$ the same as
  their direction in the catalog. This {\it semi-isotropic null} is a quick-way to check
  significance. We perform 100,000 realizations of drawing random numbers to
  assign new $RA$ and $Dec$ values for each event to construct
  $\delta\chi^2$ distributions in the same way as done with real data.

We calculate statistical significance of correlation
  in real data or $p$-value (chance probability) using frequentists'
  approach. We count the number of times we get a random data set that
  gives equal or more hits than the $N_{\rm hits}$ in real data within
  $\delta\chi^2 \le 1$ bin.  Dividing this number with the total
  number of random data sets generated (100,000) gives us the
  $p$-value. We cross-check this $p$-value by calculating the Poisson
  probability of obtaining $N_{\rm hits}$ within $\delta\chi^2 \le 1$
  bin given the corresponding average hits expected from the null
  distribution. We found the $N_{\rm hits}$ distribution in $\delta\chi^2 \le 1$ does not follow the Poisson distribution.
\section{Results and Discussions}
\begin{figure}[ht]
\begin{minipage}{16pc}
\includegraphics[width=16pc]{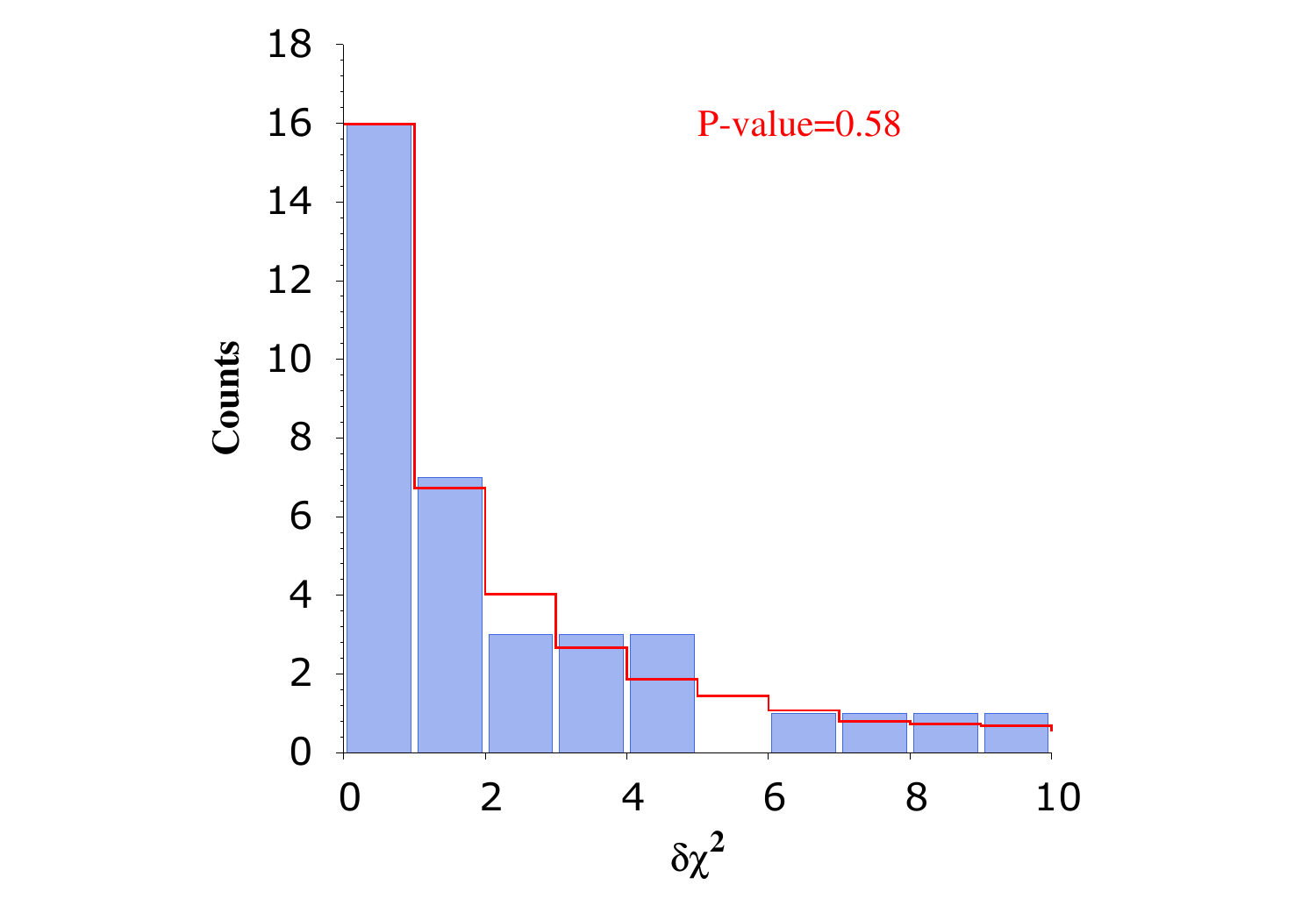}
\caption{\label{hbl}Correlation Study for all 45 HBL sources from TeVCat.}
\end{minipage}\hspace{2pc}%
\begin{minipage}{18pc}
\includegraphics[width=16pc]{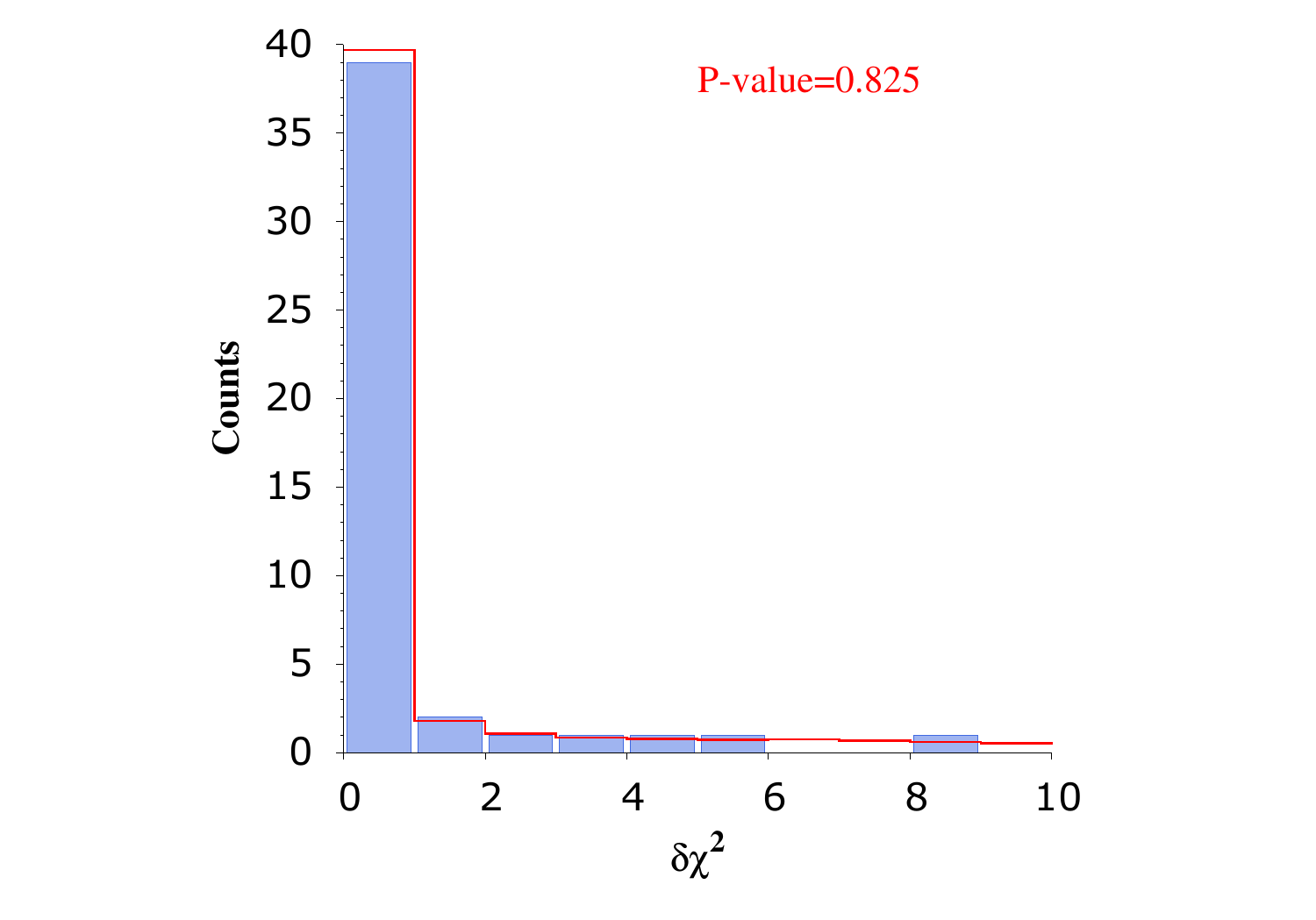}
\caption{\label{swift}Correlation Study for {\it Swift} BAT X-ray catalog sources with energy flux more than $10^{-11} {\rm{erg} \, \rm{cm^{-2}} \, \rm{sec}^ {-1}}$ also shown in ~\cite{rewin}.}
\end{minipage} 
\end{figure}

 We used all 45 HBL (high-frequency peaked BL Lacs) type source listed in TeVCat for our first  correlation study with neutrino events.  A similar correlation study was carried out in ~\cite{Sahu:2014fua} using HBLs and neutrino data. Our study showed a $p$-value 0.58 with frequentists method while with Poisson distribution probability is 0.1, with 16 neutrinos correlating with different HBLs, almost the same as the null distribution. The distribution is shown in Fig.~\ref{hbl}.
  
 {\it Swift}-BAT 70 month X-ray source included 657 sources with energy flux $10^{-11} {\rm{erg} \, \rm{cm^{-2}} \, \rm{sec}^ {-1}}$. The study of correlation with neutrino events showed a $p$-value 0.825 with 39 $N_{\rm hits}$ for the real data and nearly 40 for null distribution, as shown in Fig.~\ref{swift}.
 
\begin{figure}[h]
\begin{minipage}{16pc}
\includegraphics[width=16pc]{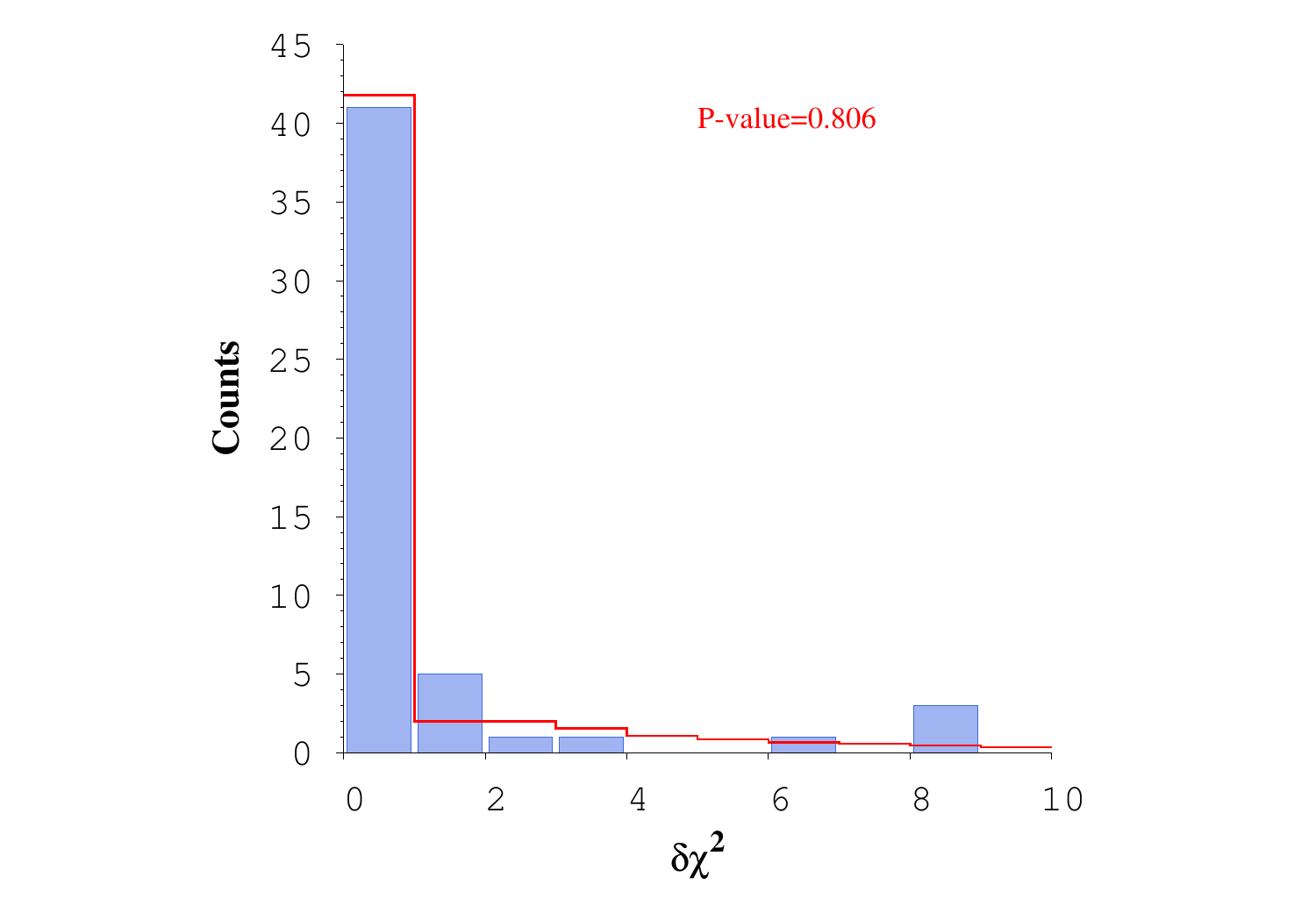}
\caption{\label{3lac_all}Correlation Study for all 1773 sources of extended 3LAC catalog also shown in ~\cite{rewin}.}
\end{minipage}\hspace{2pc}%
\begin{minipage}{16pc}
\includegraphics[width=16pc]{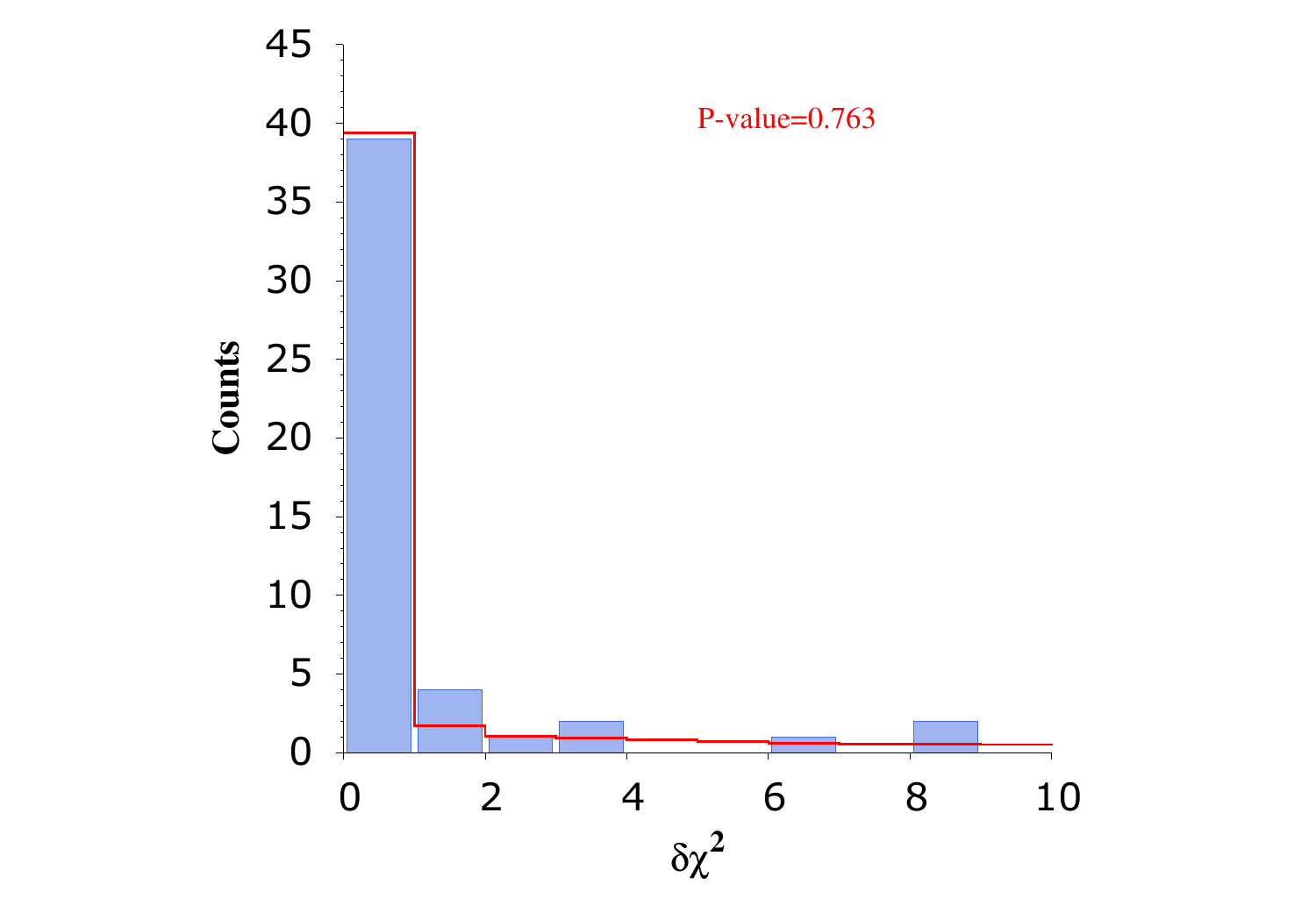}
\caption{\label{3lac_f}Correlation Study for sources from extended 3LAC catalog with energy flux $\ge$ $10^{-11} {\rm{erg} \, \rm{cm^{-2}} \, \rm{sec}^ {-1}}$.}
\end{minipage} 
\end{figure}

The correlation study of all 1773 sources in the extended 3LAC catalog gives a $p$-value 0.806 with 41 $N_{\rm hits}$ in for real data, as shown in Fig.~\ref{3lac_all}. Most of the 3LAC sources are populated in the region of energy flux $10^{-11} {\rm{erg} \, \rm{cm^{-2}} \, \rm{sec}^ {-1}}$, and the population decreases abruptly at higher flux. So, we took a set of sources with energy flux  $\ge 10^{-11} {\rm{erg} \, \rm{cm^{-2}} \, \rm{sec}^ {-1}}$. It decreased the number of sources in the set to 652, and the correlation study has a $p$-value 0.763 , with $N_{\rm hits}$ in $\delta\chi^2 \le 1$, 39, shown in Fig.~\ref{3lac_f}.

\begin{figure}[h]
\begin{minipage}{16pc}
\includegraphics[width=16pc]{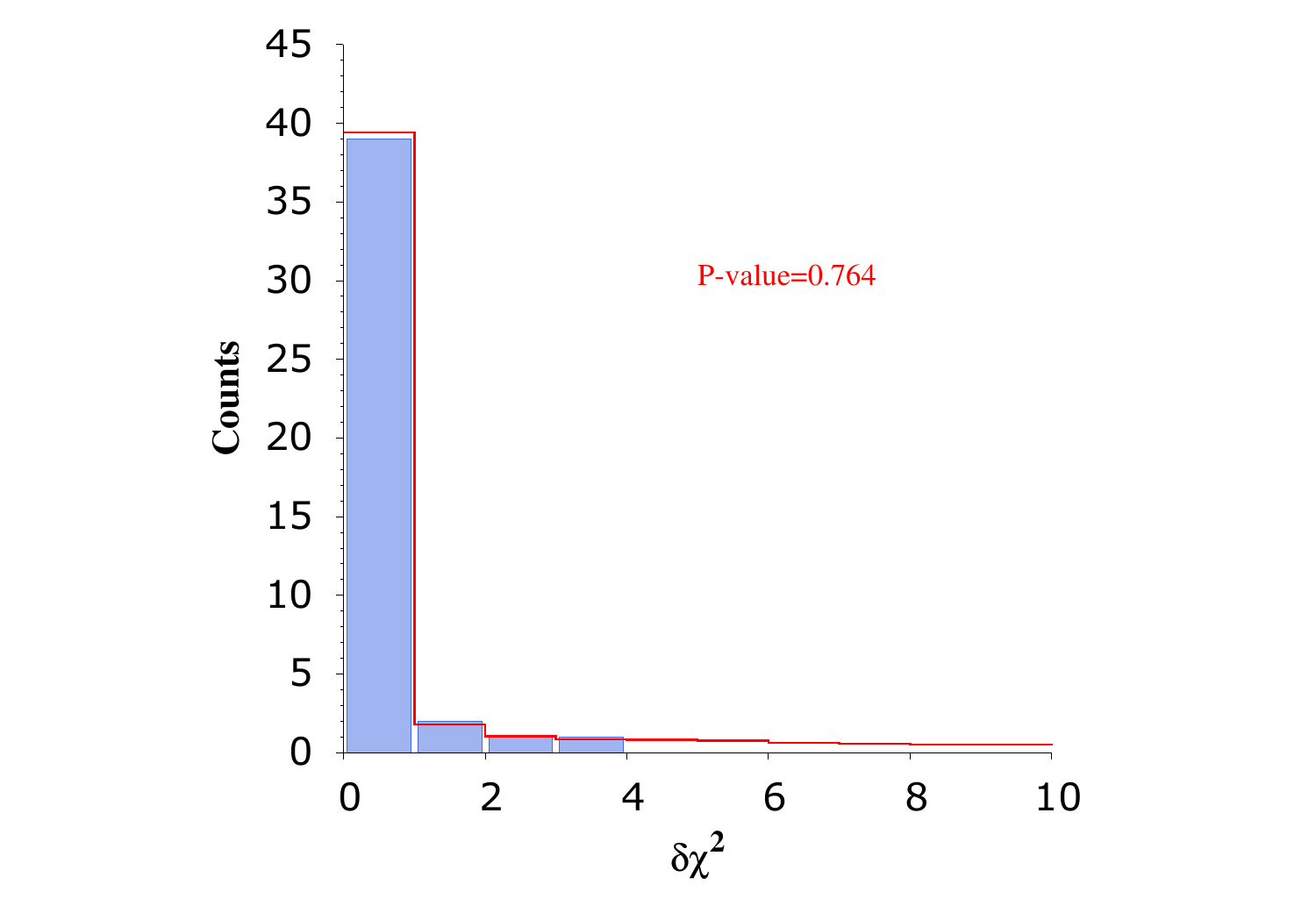}
\caption{\label{bllac}Correlation Study of BL Lac sources from extended 3LAC catalog.}
\end{minipage}\hspace{2pc}%
\begin{minipage}{16pc}
\includegraphics[width=16pc]{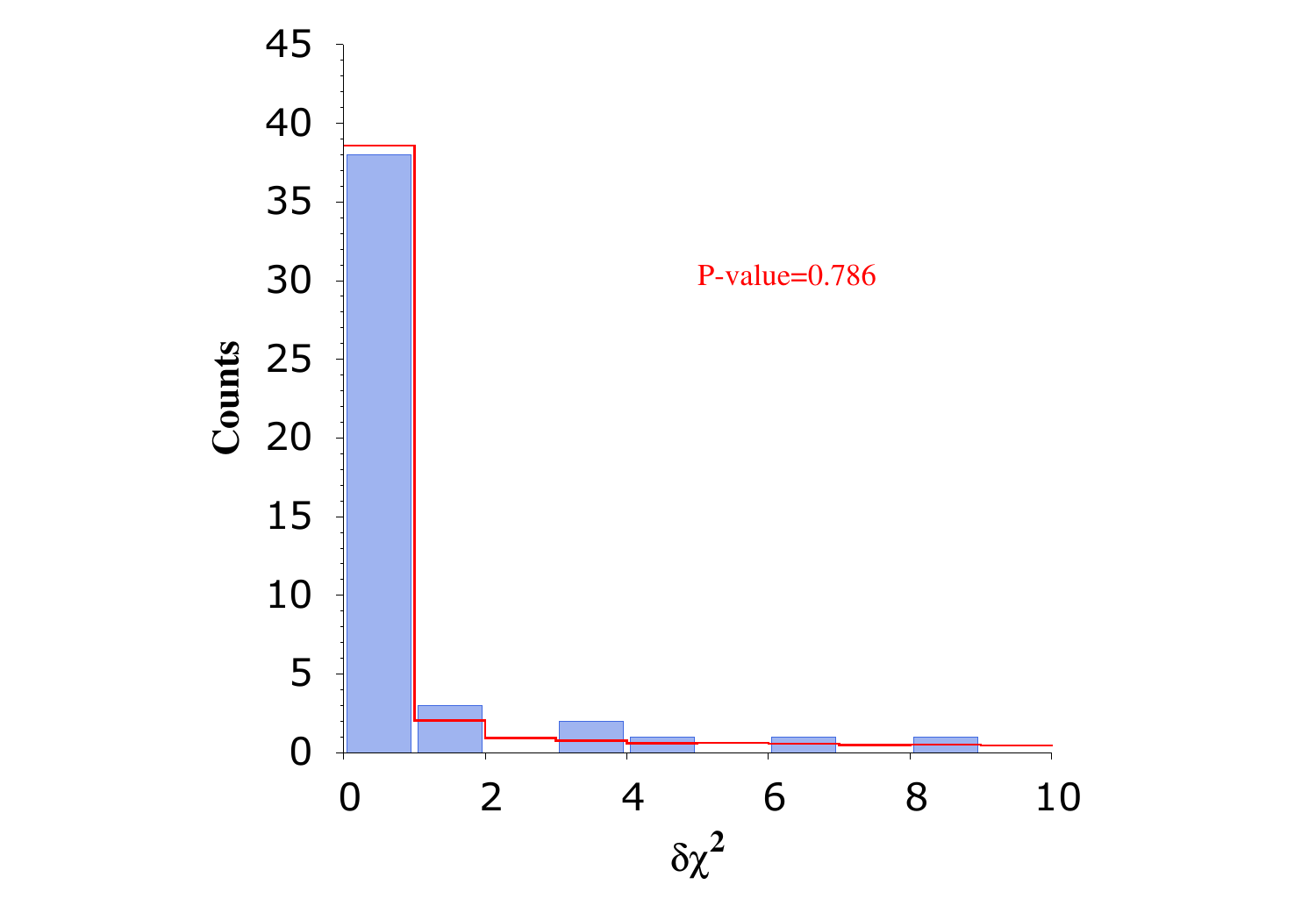}
\caption{\label{fsrq}Correlation Study of FSRQ sources from extended 3LAC catalog.}
\end{minipage} 
\end{figure}

In order to do further study for different type of sources we used the 662 BL Lac source set from the extended 3LAC catalog. The correlation $p$-value for these sources is  0.764, shown in Fig. ~\ref{bllac}. Similarly for the 491 FSRQ sources  from the extended 3LAC catalog the $p$-value is 0.784, shown in Fig.~\ref{fsrq}. For BL Lac and FSRQ sources we found 39 and 38 $N_{\rm hits}$ respectively.

The correlation study of IceCube neutrino events with different type of sources as TeVCat HBL, 3LAC BL Lac and FSRQ is done but we have not found any statistically significant result for these sets. We have also put constraints on the energy flux of 3LAC catalog and the sources observed by {\it Swift} in 70 months of its observation, and the result is not significant. However with this type of study we can discard different type of extragalactic sources for IceCube neutrino events.

\begin{table*}\centering
\begin{tabular}{|c|c|c|c|} \hline
{Catalog Name} & {Source type} & {\# of sources} & {p-value}
\\  \hline
TeVCAT              &   HBL                                                                  & 45      & 0.58    \\
$Swift$ Bat X-ray   & energy flux > $10^{-11} \,{\rm{erg} \, \rm{cm^{-2}} \, \rm{sec}^ {-1}}$             & 657     & 0.825    \\
3LAC (Extended)               & All                                                                    & 1773    & 0.806   \\
3LAC (Extended)              & energy flux > $10 ^{-11} \,{\rm{erg} \, \rm{cm^{-2}} \, \rm{sec}^ {-1}}$             & 652     & 0.763             \\
3LAC (Extended)              & BL Lac                                                                 & 662     & 0.764    \\
3LAC (Extended)              & FSRQ                                                                   & 491     & 0.786   \\
 \hline
 \end{tabular}
\caption{Results of correlation study. }
\label{tab:res}

\end{table*}

\section{Summary}
IceCube neutrino observatory has detected at least 54 neutrino events within energy 30 TeV-2 PeV. Sources for these events is still a puzzle for both particle physics and astrophysics. In our project we have tried to find correlation of the arrival direction of these events with direction of sources from TeVCat, {\it Swift} and 3LAC catalogs. In order to test correlation we have used invariant statistics, called the minimum $\delta \chi^2$, as in ~\cite{Virmani:2002xk,Moharana:2015nxa}. Out of 52 neutrino events, 16 were correlated with HBLs from TeVCat but the statistical significance of this correlation $p-$value is 0.58. Similarly we study correlation of neutrino events with sources from {\it Swift} and 3LAC having energy flux $\ge$ $10^{-11} {\rm{erg} \, \rm{cm^{-2}} \, \rm{sec}^ {-1}}$, for which we also found a poor statistical significance. The FSRQ and BL Lacs from 3LAC catalog also showed less significant statistics for the correlation study.

\end{document}